# IP over Voice-over-IP for censorship circumvention

Amir Houmansadr    Thomas Riedl    Nikita Borisov
Andrew Singer
Electrical and Computer Engineering Department
University of Illinois at Urbana-Champaign
{ahouman2,triedl2,nikita,acsinger}@illinois.edu

November 8, 2018

## Abstract

Open communication over the Internet poses a serious threat to countries with repressive regimes, leading them to develop and deploy network-based censorship mechanisms within their networks. Existing censorship circumvention systems face different difficulties in providing *unobservable* communication with their clients; this limits their *availability* and poses threats to their users. To provide the required unobservability, several recent circumvention systems suggest modifying Internet routers running outside the censored region to intercept and redirect packets to censored destinations. However, these approaches require modifications to ISP networks, and hence requires cooperation from ISP operators and/or network equipment vendors, presenting a substantial deployment challenge.

In this paper we propose a deployable and unobservable censorship-resistant infrastructure, called FreeWave. FreeWave works by modulating a client's Internet connections into acoustic signals that are carried over VoIP connections. Such VoIP connections are targeted to a server, FreeWave server, that extracts the tunneled traffic of clients and proxies them to the uncensored Internet. The use of actual VoIP connections, as opposed to traffic morphing, allows FreeWave to relay its VoIP connections through oblivious VoIP nodes, hence keeping itself unblockable from censors that perform IP address blocking. Also, the use of end-to-end encryption prevents censors from identifying FreeWave's VoIP connections using packet content filtering technologies, like deep-packet inspection.

We prototype the designed FreeWave system over the popular VoIP system of Skype. We show that FreeWave is able to reliably achieve communication bandwidths that are sufficient for web browsing, even when clients are far distanced from the FreeWave server.

# 1 Introduction

The Internet is playing an ever increasing role in connecting people from across the world, facilitating the free circulation of speech, ideas and information. This poses a serious threat to the benefits and, even, the existence of repressive regimes as it elevates their citizens' awareness and provides them a powerful media to arrange coordinated opposition movements. The recent unrest in the Middle East indicates the very strong power of the Internet in arranging nation-wide protests that, in several cases, has resulted in revolutionizing or even overthrowing repressive regimes. To response to such threats, repressive regimes make use of different technologies to restrict and monitor their citizens' access to the Internet, known as *Internet censorship*. Censorship devices, called *censors*, leverage techniques [11, 33] ranging from simple IP address blocking and DNS hijacking, to the more complicated and resource-intensive deep packet inspection (DPI) in order to enforce their restrictions and monitoring. Citizens identified non-complying with such restrictions can face consequences ranging from Internet service disruption to life-threatening punishments.

To help the censored users gain open access to the Internet different systems and technologies have been designed and developed, generally referred to as *censorship circumvention* tools. These systems are composed of computer and networking technologies and allow Internet users to evade monitoring, blocking, and tracing of their activities over the Internet. Early circumvention systems work by *proxy*ing the Internet traffic of their users to censored Internet destinations, foiling IP-address blocking and DNS hijacking performed by censors. Examples of such systems are DynaWeb [1], Anonymizer [20], and Freenet [23]. The use of advanced content filtering technologies [11, 33], i.e., deep packet inspection, by censors renders the use of simple proxies ineffective for censorship evasion. This resulted in the advent of more sophisticated circumvention systems like Ultrasurf [8] and Psiphon [4] that try to confuse content filtering tools by obfuscating their design and/or traffic patterns. Such obfuscation does not provide strong security promises for users, as investigated in recent study [16]. More recently, several circumvention systems are designed that work by hiding the circumvented traffic in regular Internet traffic [21, 27, 30, 48]. In addition to special-purpose circumvention systems, *anonymity systems* are also often used for censorship circumvention. Anonymity systems aim in hiding the browsing activity of users from other parties, a objective in common with circumvention systems. For instance, the Tor [26] anonymity system has recently adopted the use of semi-public nodes, known as Tor bridges [25], that help censored users to gain unrestricted access to the Internet anonymously. Through the rest of this paper, by circumvention systems we consider both circumvention-centric systems and the anonymity systems adopted to evade censorship.

While these circumvention systems have long been used by many users from across the world they suffer from several important shortcomings. The biggest threat to existing circumvention systems is their weak *availability* to users, i.e., they lack *unblockability*. Censors proactively look for services that help with censorship circumvention and either disrupt their operation, or block their citi-zens from accessing them. In particular, censors rigorously look for IP addresses belonging to circumvention technologies (e.g., HTTP/SOCKS proxies) and add them to IP blacklists maintained by their censoring firewalls [3,11]. As a result, citizens under repressive regimes commonly find it difficult to access operating circumvention systems. For instance the popular Tor system has frequently been/is blocked by censorship regimes [13,46], making it unavailable to users in certain countries. To improve their availability, circumvention systems have taken different approaches to make their connections *unobservable* to censors, hence unblockable. The Tor system, for instance, has deployed Tor bridges [25] that are volunteer proxies whose IP addresses are distributed among Tor users in a selective manner. This makes Tor bridges less prone to be identified by censors, as compared to Tor entry nodes, however there are several challenges in distributing their IP addresses among users [38,39]. Other systems [21,27] try to provide unobservability (hence availability) by pre-sharing some secret information with their users. This, however, is neither scalable, nor effective, as it is challenging to share secrets with a large number of real users, while keeping them secret from censors [28,35,44]. More recently, several systems are proposed that rely on support from third-parties for their operation. For instance Telex [48] and Cirripede [30] work by concealing the circumvented traffic inside regular traffic thanks to some friendly ISPs that deflect/manipulate their intercepted traffic. The real-world deployment of such circumvention systems requires collaboration of several trusted ISPs that make software and/or hardware modifications to their infrastructure; this does not seem to be realized in short-time until there are enough financial motives for the ISPs.

In this paper, we design FreeWave, a censorship circumvention system that is highly unblockable[1]. The main idea of FreeWave, as shown in Figure 1, is to tunnel Internet traffic inside non-blocked VoIP communications by modulating them into acoustic signals that are carried over VoIP connections. For a censored user to use FreeWave for circumvention, she needs to setup a VoIP account with a public VoIP provider, and also to install FreeWave's client software on her machine. Part of the FreeWave system is a FreeWave server that listens on several publicly advertised VoIP IDs for connections from its clients. To make a FreeWave connection, a user's FreeWave client software makes VoIP connections to FreeWave server's VoIP IDs. The client and server, then, tunnel the circumvented Internet traffic inside the established VoIP connections, by modulating network packets into acoustic signals carried by the established VoIP connections.

We claim that FreeWave provides strong unblockability promises since its unblockability is tied to the availability of VoIP communications. VoIP constitutes an important part of today's Internet communications [6,9,10], hence censors are not willing to block *all* VoIP communications due to several financial and political reasons. A recent report [10] shows that about one-third of U.S. businesses use VoIP solutions to reduce their telecommunications expenses, and the report predicts the VoIP penetration to reach 79% by 2013, a 50% increase compared to 2009. FreeWave's availability is tied to the availability of VoIP service: as the operation of FreeWave is not bound to a specific VoIP provider, in order to block FreeWave a censor needs to block *all* VoIP service providers. We also claim that a censor can not distinguish FreeWave-related VoIP traffic from regular VoIP traffic: the modulation of circumvented traffic inside real acoustic signals, as well as the use of encryption and coding algorithms makes FreeWave's VoIP connections *unobservable* to censors. In addition, the way FreeWave server is connected to the Internet results in getting FreeWave's VoIP connections to be *relayed* by various *oblivious VoIP peers*, preventing a censor from blocking/identifying FreeWave's VoIP connections by matching IP addresses (see Figure 1). In fact, this is the key feature distinguishing FreeWave from similar circumvention systems that work by morphing circumvented traffic to look like other protocols. For instance, SkypeMorph [40] suggests to morph (i.e., shape) a Tor client's traffic to a Tor bridge to look like an encrypted Skype connection, in order to prevent a censor from identifying the connection as a Tor connection and, hence, detecting the IP address of the morphing Tor bridge. This, however, does not mitigate the big issue of distributing bridge IP address (and Skype IDs for SkypeMorph), so censors can eventually identify IP addresses belonging to the morphing Tor bridges (e.g., by pretending to be real clients), hence block the bridges and/or punish the identified clients. In FreeWave, however, the VoIP connections carrying circumvented traffic are established between clients and some oblivious VoIP peers, so even a censor's knowledge of FreeWave server's IP address can not help him in identifying/blocking FreeWave's VoIP connections (this is in contrast to SkypeMorph where morphed VoIP connections are between clients and the Tor bridges).

---

[1]We use terms *unblockability* and *availability* interchangeably.

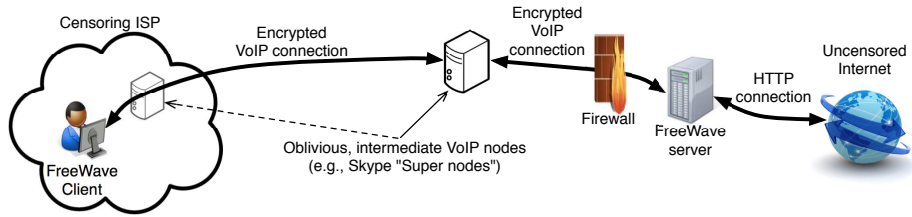

Figure 1: The main architecture of FreeWave.

We implement a prototype of FreeWave system over Skype and measure its performance. Our evaluations show that FreeWave provides connection bit rates that are convenient for regular web browsing. We also validate that FreeWave can be used by clients that are geographically far away from the FreeWave server, with a reasonably lower connection throughput.

The rest of this paper is organized as follows; in Section 2 we review related research. In Section 3 we review our threat model and the goals in designing our circumvention system. We describe the design of our proposed circumvention system, FreeWave, in Section 4, and Section 5 discusses our design details. In Section 6, we discuss the features of our designed circumvention system.

Section 7 describes the FreeWave's MoDem block, an essential component of FreeWave that modulates Internet traffic into acoustic signals. We describe our prototype implementation of FreeWave in Section 8, along with the evaluation results. Finally, the paper is concluded in Section 9.

## 2  Related work

Since the epoch of Internet censorship, different entities have designed and built systems to combat censorship. This has resulted in an arms-race between censors and circumvention systems, a game with no ultimate winner as of now. Early circumventions systems are based on setting up a *proxy* [24] server outside the censorship territories, trying to evade the simple IP address blocking and DNS hijacking techniques enforced by pioneer censorship systems. Examples of such proxy-based circumvention tools are DynaWeb [1], Anonymizer [20], and Freenet [23].

Proxy-based circumvention tools lost their capabilities with the advent of more sophisticated censorship technologies, like deep-packet inspection [11, 33]. Deep-packet inspection analyzes packet contents looking for patterns indicating the circumvention of the enforced censorship. This has pushed the circumvention tools to correspondingly sophisticate their techniques in order to remain available to users. In particular, several circumvention system base their availability on being unobservable from censors. Infranet [27], for example, works by sharing some secret information with its clients, which are then used by the clients to establish unobservable connections with Infranet servers. Collage [21] is another system that, similarly, bases its unobservability on sharing secrets with its clients. More specifically, a Collage client and the Collage server secretly agree on some user-generated content sharing websites, e.g., flickr.com, and use steganography to communicate through these websites. The main challenge for these systems is to share secret information with a large set of users such that censors do not get hands on the shared secrets; this is a challenging problem to solve as discussed in several research [28, 35, 44]. Sharing secret information with users has also been adopted by the popular Tor anonymity network [26] to help with censorship circumvention. The secret information here are the IP address of several volunteer nodes, Tor bridges [25], that proxy the connections of Tor clients to the Tor network. This suffers from the same limitation as censors can pretend to be real Tor users and gradually identify a large fraction of Tor bridges [38, 39].

More recently, several research propose to build circumvention into the Internet infrastructure [30, 31, 48]. Being built into the Internet infrastructure makes such circumvention highly unobservable: a client's covert communication with a censored destination appears to the censor to be a benign traffic with a non-prohibited destination. Telex [48], Cirripede [30] and Decoy Routing [31] are example designs using such infrastructure-embedded approach. Decoy Routing needs to share secrets with its clients using out-of-band channels, whereas Telex and Cirripede share the secret information needed to initialize their connections using covert channels inside Internet traffic. Cirripede uses an additional client registration stage that provides some advantages and limitations as compared to Telex and Decoy routing systems. Even though these systems are a large step forward in providing unobservable censorship circumvention their practical deployment is not trivial as they need to be deployed by a number of real-world ISPs that will make software/hardware modifications to their network infrastructures, something that needs enough financial motivations for ISPs to have them deploy it.

Another research trend uses traffic obfuscation as a method to make circumvented traffic unobservable. Appelbaum et al. propose a platform that allows one to build protocol-level obfuscation plugins for Tor, called *pluggable transports* [17]. These plugins obfuscate a Tor client's traffic to Tor bridges by hiding it into another protocol that is allowed by censors. There are two pluggable transports available; obfsproxy [36] obfuscates Tor traffic by passing it through a stream cipher. Mohajeri et al. [40] show that obfsproxy generates traffic patterns that are detectable by powerful censors. They, consequently, propose SkypeMorph [40] that morphs a Tor client's traffic into Skype video calls, in order to make it undetectable by deep-packet inspection. We argue that even though this prevents a censor from detecting Tor bridges through traffic analysis, the censors can still pretend to be real Tor users and gradually gain the IP addresses of Tor bridges, hence blocking their connections. The authors of SkypeMorph propose frequent change of the detected IP addresses, as well as the detected Skype IDs in response to this. As another possible issue, in SkypeMorph clients need to send key initialization information to tor bridges through Skype text messages. As such messages have a known format they can be detected and blocked by Skype; in particular, since 2007 Skype provides a special version of its client software to Chinese users that monitors and blocks the text messages sent by Chinese users [12, 14, 15, 22]. CensorSpoofer [45] is another recent research that aims in providing unobservable circumvention by morphing circumvented traffic into spoofed VoIP connections. A common security concern with morphing approaches is that they do not provide a provable undistinguishability between a morphed traffic and the regular traffic; this may result in finding effective statistical tests, in the future, that are able to detect currently-undetectable morphing techniques, e.g., [40] mentions the possibility of detecting Traffic Morphing [47] by analyzing inter-packet delays. As another similar circumvention system, TranSteg [37] re-encodes the call stream in a VoIP connection using a different codec in order to free a portion of VoIP packet payloads, which are then used to send a low-bandwidth circumvented traffic.

# 3 Preliminaries

## 3.1 Threat model

We assume that a FreeWave user, namely FreeWave client, is connecting to the Internet through a censoring ISP, i.e., an ISP that is controlled and regulated by a censorship authority. Based on the regulations of the censoring ISP, its users are not allowed to connect to certain Internet destinations, called the *censored destinations*. The users are also prohibited from using censorship circumvention technologies that would help them to evade the censoring regulations. The censoring ISP uses a set of advanced technologies to enforce its censoring regulations, including IP address blocking, DNS hijacking, and deep packet inspection. The censoring ISP also monitors its users' network traffic to identify and block any usage of censorship circumvention tools; traffic analysis can be used by the censor as a powerful technique for this purpose.

We assume that the censoring ISP enforces its regulations such that it does not compromise the *usability* of the Internet, due to different political and economical reasons. In other words, the enforced censorship does not disable/disrupt key services of the Internet. In particular, we consider VoIP as a key Internet service in today's Internet [6,7,9], and we assume that, even though a censor may block certain VoIP providers, it will not block *all* VoIP services. VoIP constitutes a key part in the design of FreeWave.

## 3.2 System goals

We consider the following as the key features in designing and evaluating a censorship circumvention system. Later in Section 6.1, we discuss these features for the FreeWave circumvention system proposed in the consequent sections and compare FreeWave with related work.

**Unblockability:** The main goal of a censorship circumvention system is to help censored users gain access to censored Internet destinations. As a result, the most trivial property of a circumvention system is being accessible by censored users, i.e., it should not be unblockable by censors.

**Unobservability:** Unobservability is to hide users' utilization of a circumvention system from censorship authorities, which is a challenging feature to achieve due to the recent advances in censorship technologies [11]. The importance of unobservability is two-fold; first, an observable circumvention can jeopardize the safety of a user who has been caught by the censor while using the circumvention system; and, second, a weak unobservability commonly results in a weak unblockability, as it allows censors to more easily identify, hence block, traffic generated by the circumvention system.

**Security:** Several security considerations should be made once analyzing a circumvention system. These considerations include users' anonymity, security and privacy from different entities including the censors, the utilized circumvention system, and other third-parties.

**Deployment feasibility:** An important feature of a circumvention system is the amount of resources (e.g., hardware, network bandwidth, etc.) required for it to be deployed in real-world. A circumvention system is also desired to have few dependencies on other systems and entities in order to make it more reliable, secure, and cost-effective.

**Quality of service:** A key feature in making a circumvention system popular in practice is the quality of service provided by it in establishing circumvented connections. Two important factors are the connection bandwidth, and the browsing latency.

## 4 FreeWave scheme

In this section, we describe the design of FreeWave censorship circumvention. Figure 1 shows the main architecture of FreeWave. In order to get connected through FreeWave, a user installs a *FreeWave client* on her machine, which can be obtained from an out-of-band channel, similar to other circumvention systems. The user sets up the installed FreeWave client by entering her own VoIP ID and also the publicly known VoIP ID of FreeWave server. Once the FreeWave client starts up, it makes a VoIP audio/video call to FreeWave server's VoIP ID. As discussed in Section 5.2, the FreeWave server is configured in a way that VoIP connections initiated by clients are relayed through various *oblivious VoIP peer*s, e.g., Skype supernodes; this is a key security feature of FreeWave as it prevents a censor from blocking FreeWave's VoIP connections using IP address blocking. Also, since FreeWave's VoIP connections are end-to-end encrypted, a censor will not be able to identify FreeWave's VoIP connections by analyzing traffic contents, e.g., by looking for the VoIP IDs. Using the established VoIP connection, a FreeWave client circumvents censorship by modulating its user's Internet traffic into acoustic signals that are carried over by such VoIP connections. FreeWave server demodulates a client's Internet traffic from the received acoustic signals, and proxies the demodulated traffic to the requested Internet destinations.

Next, we introduce the main components used in FreeWave and, then, describe how these components are used in the design of FreeWave's client and server.

### 4.1 Components of FreeWave

In this section, we introduce the main elements used in the design of FreeWave client and server software. The first three components are used by both FreeWave client and FreeWave server, while the fourth element is only used by FreeWave server.

**VoIP client** A Voice-over-IP (VoIP) client software that allows VoIP users to connect to one (or more) specific VoIP service(s). In Section 5.2, we discuss the choices of the VoIP service being used by FreeWave.

**Virtual sound card (VSC)** A virtual sound card is a software application that uses a physical sound card installed on a machine to generate one (or more) isolated, virtual sound card interfaces on that machine. A virtual sound card interface can be used by any application running on the host machine exactly the same way a physical sound card is utilized. Also, the audio captured or played by a virtual sound card does not interfere with that of other physical/virtual sound interfaces installed on the same machine. We use virtual sound cards in the design of FreeWave to isolate the audio signals generated by FreeWave from the audio corresponding to other applications.

**MoDem** FreeWave client and server software use a modulator/demodular (MoDem) application that translates network traffic into acoustic signals and vice versa. This allows FreeWave to tunnel the network traffic of its clients over VoIP connections by modulating them into acoustic signals. We provide a detailed description of the design of our MoDem in Section 7.

**Proxy** FreeWave server uses an ordinary proxy server application that proxies the network traffic of FreeWave clients, received over VoIP connections, to their final Internet destinations. Two popular choices for FreeWave's proxy are the HTTP proxy [41] and the SOCKS proxy [34] protocols; a SOCKS proxy supports proxying of a wide range of IP protocols, while an HTTP proxy only supports proxying of HTTP/HTTPS traffic, but it can perform HTTP-layer optimizations like pre-fetching of web contents. Several proxy solutions support both protocols.

## 4.2  Client design

The FreeWave client software, installed by a FreeWave user, is consisted of the three main components listed above: a VoIP client application, a virtual sound card (VSC), and the MoDem application. Figure 2 shows the block diagram of the FreeWave client design. MoDem transforms the data of the network connections sent by the web browser into acoustic signals and sends them over to the VSC component. The FreeWave MoDem also listens on the VSC sound card to receive specially formatted acoustic signals that carry modulated Internet traffic; MoDem extracts the modulated Internet traffic from such acoustic signals and sends them to the web browser. In fact, client's web browser uses the MoDem component as a web proxy, i.e., the listening port of MoDem is entered in the HTTP/SOCKS proxy settings of the browser.

The VSC sound card acts as a bridge between MoDem and the VoIP client component, i.e., it transfers audio signals between them. More specifically, the VoIP client is set up to use the VSC sound card as its *speaker* and *microphone* devices (VoIP applications allow a user to specify the (physical/virtual) sound cards to be used with them). This provides MoDem and the VoIP client to exchange audio signals that contain the modulated network traffic, isolated from the audio generated/recorded by other applications on the client machine.

For the FreeWave client to connect to a particular FreeWave server it *only* needs to know the VoIP ID belonging to that FreeWave server, but not the IP address of the FreeWave server. Every time the user starts up the FreeWave client application on her machine the VoIP application of FreeWave client initiates an audio/video VoIP call to the known VoIP ID of the FreeWave server.

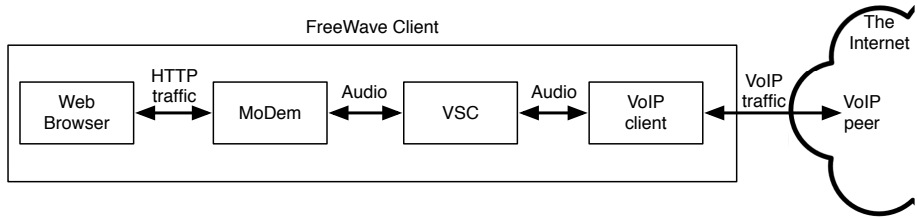

Figure 2: The main components of FreeWave client.

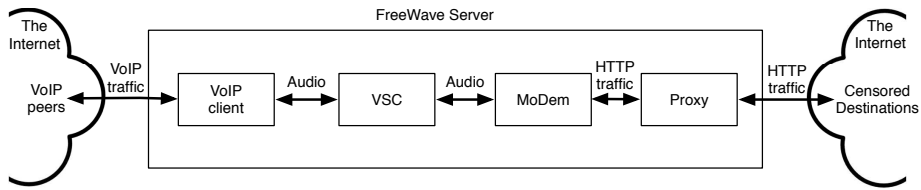

Figure 3: The main components of FreeWave server.

### 4.3 Server design

Figure 3 shows the design of FreeWave server, which consists of four main elements. FreeWave server uses a VoIP client application to connect with its clients through VoIP connections. FreeWave server chooses one or more VoIP IDs, which are provided to its clients, e.g., through public advertisement.

The VOIP client of FreeWave server uses one (or more) virtual sound cards (VSC) as its *speaker* and *microphone* devices. The number of VSCs used by the server depends on the deployment scenario, as discussed in Section 5.1. The VSC(s) are also being used by a MoDem component, that transforms acoustic signals into network traffic and vice versa. More specifically, MoDem extracts the Internet traffic modulated by FreeWave clients into audio signals from VoIP connections and forwards them to the last element of the FreeWave server, FreeWave *proxy*. The MoDem also modulates Internet traffic received from the proxy component into acoustic signals and sends them to the VoIP client software through the VSC interface. The FreeWave proxy is a regular network proxy, e.g., an HTTP proxy, that is used by FreeWave server to connect FreeWave clients to the Internet in an unrestricted manner. As mentioned above in Section 4.2, the web browser of a FreeWave client targets its traffic to a network proxy; such proxied traffic is received by FreeWave server's proxy server (through the VoIP connections, as described) and is processed as required.

# 5 Other design details

## 5.1 Deployment scenarios

The FreeWave system proposed in this paper can be deployed by "good" entities that run FreeWave servers to help censored users gain an uncensored access to the Internet. We consider the following scenarios for a real-world deployment of FreeWave. In Section 6.2, we discuss the security considerations for each of these scenarios.

- Personal deployment: A person having an uncensored access to the Internet can set up a personal FreeWave server on her personal machine, trying to *anonymously* help censored users evade censorship. Such a person can, then, advertise her VoIP ID (associated with her FreeWave server) publicly (e.g., through social networks) and anyone learning this ID would be able to connect to the Internet by running a FreeWave client software. To save her bandwidth, such a person can configure its FreeWave server to enforce restrictions on the quality of service provided to clients.

- Central VoIP-center: FreeWave service can be deployed and maintained by a central authority, e.g., a for-profit or non-profit organization. The deploying organization can build and run FreeWave servers that are a able to serve a large number of FreeWave clients. To be able to support large numbers of clients, the used FreeWave servers should deploy several physical/virtual sound cards. Also, by generating VoIP IDs on several, different VoIP service providers such central FreeWave system will be able to service FreeWave clients with different VoIP preferences. Such a central deployment of FreeWave can operate for commercial profit, e.g., by charging clients for using the system, or can be established as a non-profit systems, e.g., being funded by NGOs or pro-freedom governments.

- Central phone-center: As an alternative approach, FreeWave can be deployed using an automated telephone center. More specifically, instead of VoIP IDs, the FreeWave server will publicize several phone numbers, which are used by clients to connect to the FreeWave server. FreeWave users need to use the exact same FreeWave client software, except that instead of making VoIP calls to FreeWave server's VoIP IDs they will make VoIP calls to FreeWave server's phone numbers. Compared to the "central VoIP-center" approach, this has the big advantage that clients can arbitrarily choose any VoIP service provider for the client software, while in the "central service" design users need to choose from the VoIP systems supported by FreeWave server (though a powerful FreeWave server can support many VoIP systems).

- Distributed service: FreeWave service can also be deployed in a distributed architecture, similar to that of the popular anonymity network of Tor [26]. More specifically, FreeWave server can be consisted of a number of volunteer computers that run instances of FreeWave server software on their machines. A central authority can manage the addition of new volunteer nodes to the system and also the advertisement (or distribution) of their VoIP IDs to the clients.

## 5.2 The choice of VoIP system

There are numerous freeware and paid *VoIP service provider*s that can be utilized by the FreeWave system, e.g., Skype[2], Vonage[3], iCal[4], etc. Each VoIP service provider provides an application software for its clients; also, there are software that support connection to different VoIP providers, e.g., PhonerLite[5]. In this section, we mention possible VoIP service provider candidates to be used by FreeWave.

### 5.2.1 Skype

Skype is a peer-to-peer VoIP system that provides voice calls, instant messaging, and video calls to its clients over the Internet. Skype is one of the most popular VoIP service providers with over 663 million users[6] as of September 2011.

Skype uses an undisclosed proprietary design, which has been analyzed in several research [18, 29, 49]. These studies find that Skype uses a peer-to-peer overlay network with the Skype users as its nodes. There are two types of nodes on Skype: *ordinary nodes*, and *super nodes (SN)*. Any Skype client with a public IP address, having sufficient CPU, memory, and network bandwidth is considered a super nodes, and all the other nodes are considered as ordinary nodes. In addition, Skype uses a central *login server* that keeps users' login credentials and is used by Skype users to register into Skype's overlay network. Apart from the login server all Skype communications work in a peer-to-peer manner, including the user search queries and online/offline user information.

A key feature that makes Skype an ideal choice for FreeWave is its peer-to-peer network. In particular, a VoIP connection made towards an ordinary node in Skype gets relayed by at least one oblivious Skype super node [18, 29]. More specifically, each ordinary node maintains a *host cache (HC)* table that keeps a list of reachable super nodes, which are used by the ordinary node to make Skype connections. We use this feature to protect FreeWave from being blocked by censors: by having our FreeWave server to act as an ordinary node the VoIP connections from FreeWave clients to the FreeWave server will be relayed by alternative super nodes, hence censors will not be able to enforce IP address blocking. We discuss this further in Section 6.2. Note that a FreeWave server can have its VoIP connections be relayed by arbitrarily various super nodes by frequently *flush*ing its HC table [18].

Based on the criteria mentioned for a super node, an easy way to be considered an ordinary node by Skype is to sit behind NAT or a firewall [29] (since a super node must have a static public IP address). As a result, in order to have a FreeWave server to be considered as a Skype ordinary node, it can simply be protected by a firewall or can be put behind a NAT with dynamic address allocations. As another interesting feature for FreeWave, all Skype connections are secured by end-to-end encryption [18, 29].

---

[2] http://www.skype.com
[3] http://www.vonage.com
[4] http://www.icall.com/
[5] http://www.phonerlite.de/index_en.htm
[6] http://www.telecompaper.com/news/skype-grows-fy-revenues-20-reaches-663-mln-users

### 5.2.2 SIP-based VoIP

Session initiation protocol [42] (SIP) is a light-wight, popular signaling protocol and is widely used by VoIP providers, e.g., SFLphone[7], Zfone[8], and Blink[9], to establish calls between clients. A SIP-based VoIP system consists of three main elements [42]: 1) *user agents* that try to establish SIP connections on behalf of users, 2) a *location service* that is a database keeping information about the users, and 3) a number of *servers* that help users in establishing SIP connections. In particular, there are two types of SIP servers; *registrar* servers receive registration requests sent by user agents and update the location service database. The second type of SIP servers are the *proxy* servers that receive SIP requests from user agents and other SIP proxies and help in establishing the SIP connections.

Once a SIP connection is established between two user agents a media delivery protocol is used to transfer the media among the users. In particular, most of the SIP-based VoIP systems use the Real-time Transport Protocol [43] (RTP) to exchange audio data, and use the Real-Time Transport Control Protocol [43] (RTCP) protocol to control the established RTP connections. User agents in SIP-based VoIP system are allowed to use an encryption-enabled version of RTP, called Secure Real-time Transport Protocol [19] (SRTP), in order to secure their VoIP calls. Note that the encryption supported by SRTP is done end-to-end and the VoIP servers are not required to support encryption. We mandate the SIP-based design of FreeWave to use SRTP for media transfer.

Similar to the case of Skype, if a user agent is behind a NAT or a firewall it will use an intermediate node to establish its VoIP connections. In particular, two popular techniques used by VoIP service providers to bypass NAT and firewalls are *session border controller* (SBC) [5] and *RTP bridge server*s [2]. As in the case of the Skype-based FreeWave, putting a FreeWave server behind a firewall masks its IP address from censors, as the VoIP calls to it will be relayed through oblivious intermediate nodes.

## 6 Discussions

### 6.1 Evalution of the design goals

In Section 3.2, we listed several features that we consider in designing an effective circumvention system. Here, we discuss the extent to which our proposed

---

[7] http://sflphone.org/
[8] http://zfoneproject.com/
[9] http://icanblink.com/

system, FreeWave, achieves such requirements.

**Unblockability:** In order to use FreeWave, a client only needs to know the VoIP ID of the FreeWave server, i.e., `server-id`, but no other secret/public information like the server's IP address. `server-id` is distributed in a public manner to the users, so we assume that it is also known to censors. Considering the use of encrypted VoIP connections by FreeWave, this public knowledge of `server-id` does not allow censors to identify (and block) the VoIP connections to FreeWave server. In addition, a censor will not be able to identify FreeWave's VoIP connections from their IP addresses since, as discussed in Section 5.2, the encrypted VoIP connections to the FreeWave server are relayed through oblivious, intermediate nodes. For instance, in Skype-based FreeWave the VoIP connections to the FreeWave server are relayed by oblivious Skype super nodes. Also, FreeWave server is not mapped to a particular set of super nodes, i.e., its VoIP connections are relayed through a varying set of super nodes. Another point in making FreeWave unblockable is that it does not depend on a particular VoIP system, and can use any VoIP provider for its connections. As a result, in order to block FreeWave, censors will need to block *all* VoIP services, which is very unlikely due to several political and economical considerations.

Note that unblockability is a serious challenge with many existing circumvention systems, as the information advertised for their utilization can be used by censors to block them. For example, the Tor [26] system requires its clients to connect to a public set of its IP addresses, which can be IP-filtered by censors. More recently, Tor has adopted the use of Tor *bridges* [25], which are volunteer proxies with semi-public IP addresses. Unfortunately, there are different challenges [38, 39, 44] in distributing the IP addresses of Tor bridges only to real clients, but not to the censors.

**Unobservability:** The arguments made above for FreeWave's unblockability can also be used to justify its unobservability. As mentioned above, even though FreeWave server's VoIP ID (`server-id`) is assumed to be known to censors, the end-to-end encryption of VoIP connections prevents a censor from observing users making VoIP connections to `server-id`. In addition, VoIP relays sitting between FreeWave clients and a FreeWave server, e.g., Skype super nodes, prevent identification of FreeWave connections by matching IP addresses.

**Security:** We devote Section 6.2 to discussing the security of FreeWave.

**Deployment feasibility:** The real-world deployment of FreeWave does not rely on other entities. This is in contrast to some recent designs that need collaboration from third parties for their operation. For instance, Infranet [27] requires support from some web destinations that host the circumvention servers. As another example, several recent systems [30, 31, 48] rely on collaboration from friendly ISPs for their operation.

**Quality of service:** In Section 8, we discus the connection performance provided by our prototype implementation of FreeWave. Our results show that FreeWave provides reliable connections that are good for normal web browsing.

## 6.2 Security analysis

In this section, we discuss the security of FreeWave clients to threats imposed by different entities.

### 6.2.1 Security from censors

The end-to-end encryption of VoIP connections protects the confidentiality of the data being sent by FreeWave clients from a monitoring censor, even if the censor is able to identify VoIP connections targeted to FreeWave. Such end-to-end encryption also ensures the web browsing privacy of FreeWave clients. As mentioned in Section 5.2, Skype calls are encrypted end-to-end, and, SIP-based VoIPs also provide end-to-end encryption using the SRTP protocol.

Even though FreeWave uses encrypted VoIP connections a censor may still try to identify FreeWave-generated VoIP connections by performing traffic analysis, i.e., by analyzing communication patterns. Since FreeWave makes *real* VoIP connections, traffic patterns of FreeWave's VoIP connections will be indistinguishable from those of typical VoIP traffic. In order to better disguise traffic patterns, a FreeWave client can make video calls, instead of voice-only calls, and use the audio channel of the video calls for tunneling FreeWave traffic. This can better hide the traffic pattern of FreeWave VoIP connections, as discussed in [40].

### 6.2.2 Security from FreeWave servers

A FreeWave server only knows the VoIP IDs of its client, but not their IP addresses since the VoIP connections are being relayed through intermediate VoIP nodes, as described before. As a result, unless the VoIP service is colluding with a FreeWave server, the FreeWave server will not be able to link VoIP IDs to IP addresses.

In the basic design of FreeWave mentioned above a FreeWave server can observe the traffic contents the browsing activity corresponding to a client VoIP ID, since the tunneled traffic is not always encrypted. However, a client can easily ensure security and privacy from the server by using an extra layer of encryption. For instance, a client can use FreeWave to get connected to an anonymity system like Anonymizer [20], and then use the tunnelled connection with this anonymity system to browse the Internet. This secures this client's traffic from the FreeWave server, as well as makes it confidential. Note that considering the fact that FreeWave clients are always anonymized from FreeWave servers, clients may opt not to use such an additional protection for low-sensitivity activities like web browsing.

### 6.2.3 Security from VoIP providers

The VoIP connections between FreeWave clients and servers are encrypted end-to-end; as a result, the VoIP provider will not be able to observe neither the data being communicated, nor the web destinations being browsed through such connections. However, the VoIP service provider is able to identify VoIP IDs that have made VoIP calls to FreeWave server. As a result, in order to ensure its unobservability FreeWave needs to use VoIP providers that are not colluding with censors. Note that FreeWave does not rely on a particular VoIP system and any VoIP provider can be used for its operation.

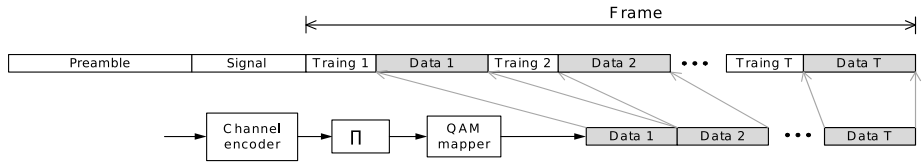

Figure 4: Frame structure

# 7  FreeWave MoDem

The MoDem component is an essential component of both FreeWave client and FreeWave server, which translates Internet traffic into acoustic signals and vice versa. MoDem consists of a transmitter and a receiver. MoDem's transmitter modulates data (IP bits) into acoustic signals, and MoDem's receiver demodulates data from a received acoustic signal. In the following, we describe the design of MoDem's transmitter and receiver.

## 7.1  Transmitter description

The transmitter structure considered here is based on bit-interleaved coded modulation (BICM). First, the information bits $\{a_i\}$ are encoded by a rate $R_c$ channel encoder, producing the coded bit sequence $\{b_i\}$. The sequence $\{b_i\}$ is permuted using a random interleaver and the resulting sequence is then partitioned into length $Q$ subsequences $\boldsymbol{c}_n = [c_n^1, \ldots, c_n^Q]$. Each such subsequence is finally mapped to a $2^Q$-ary quadrature amplitude modulation (QAM) symbol.

Data transmission is carried out frame-by-frame. As shown in Figure 4, a frame consists of $T$ sets of the training symbols followed by the data symbols. A block of known preamble symbols as well as a block of signal symbols are inserted in front of every frame. The preamble symbols are needed for synchronization and receiver initialization purposes. The signal symbols tell the receiver what modulation and coding was used for the data symbols. The training symbols are needed to adapt the equalizer to the time-varying channel. We enumerate the symbols in this frame by $x_n$.

The symbol sequence $x_n$ is mapped to a waveform $x(t) : \mathbb{R} \to \mathbb{C}$

$$x(t) = \sum_l x_l p(t - lT) \qquad (1)$$

by use of a basic pulse $p(t)$ time shifted by multiples of the symbol period $T$.

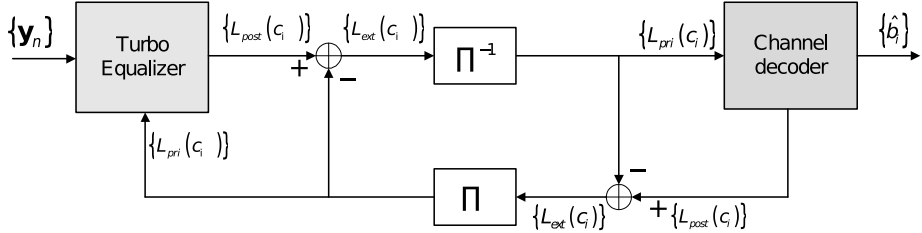

Figure 5: Block diagram of the receiver architecture. $\Pi$ and $\Pi^{-1}$ denote interleaving and deinterleaving operations, respectively.

This signal is then modulated to passband

$$x_{PB}(t) = 2\Re\{x(t)e^{2\pi i f_C t}\} \qquad (2)$$

at carrier frequency $f_C$ and sent over the Skype channel.

### 7.2 Receiver description

We assume the passband channel effect to be linear. An equivalent baseband channel representation is obtained by demodulating and low-pass filtering the received signal $r(t)$. After synchronization and symbol-spaced sampling, we obtain the equivalent discrete baseband channel observations $r_n$.

$$r_n = \sum_{k=-K_f}^{K_p} h_{n,k} x_{n-k} + w_n \qquad (3)$$

where $n$ and $k$ are time and delay indices, respectively and $w_n$ is a complex white Gaussian noise process. The channel length is assumed to be at most $K_f + K_p + 1$, where $K_f$ is the length of the precursor and $K_p$ is that of the postcursor response. We explicitly allow for the channel the gains $h_{n,k}$ to change in time $n$.

A block diagram of the used Turbo Equalizer system is shown in Figure 5. The equalizer produces extrinsic log-likelihood ratios (LLRs) on the coded bits using the received vector $y_n$ and the a priori LLRs on them [32]. Specifically, the extrinsic LLR $L_{\text{ext}}(c_n^q)$ can be obtained from $L_{\text{ext}}(c_n^q) = L_{\text{post}}(c_n^q) - L_{\text{pri}}(c_n^q)$, where $L_{\text{pri}}(c_n^q)$ is the *a priori* LLR defined as $L_{\text{pri}}(c_n^q) = \ln Pr(c_n^q = 1) - \ln Pr(c_n^q = 0)$ and $L_{\text{post}}(c_n^q)$ is the a posteriori LLR defined as $L_{\text{post}}(c_n^q) = \ln Pr(c_n^q = 1|\text{observation}) - \ln Pr(c_n^q = 0|\text{observation})$. Note that the observation can be any function of $y_n$ and the *a priori* LLRs are obtained from the channel decoder. The extrinsic LLRs produced in the equalizer are delivered to the channel decoder through the deinterleaver. They are used as *a priori* LLRs to produce the additional extrinsic LLRs in the channel decoder. Finally, the extrinsic LLRs obtained by the channel decoder are fed back to the equalizer to aid the equalization task. These steps complete one cycle of turbo iteration and are repeated until a suitably chosen convergence criterion is achieved.

# 8 Implementation and results

We build a prototype implementation of FreeWave over Skype. We use MATLAB7 to build our MoDem component, and we use Virtual Audio Card [10] as our virtual sound card (VSC) software. We also use the Skype client software provided by Skype Inc. as our VoIP client component. Our MoDem software, as well as the Skype client are set up to use the Virtual Audio Card as their audio interface. We build FreeWave client and a FreeWave server, using the components mentioned above. In order to emulate a real-world experience, i.e., a long distance between FreeWave client and FreeWave server, we connect our FreeWave client to the Internet though a VPN connection. In particular we use the SecurityKISS VPN solution that allows us to pick VPN servers located in different geographical locations around the world. In particular, in our experiments we use VPN servers that are located in Poland and USA. Note that this identifies the location of our FreeWave client, and our FreeWave server is located within USA (near Indianapolis).

**MoDem Specifications:** Our evaluations show that the data rates that can be achieved with our system clearly depend on the quality of the Skype connection. The minimum bandwidth required for a voice call is 30 kbps for both upload and download speed, according to Skype. The pulse function of $p(t)$ is a square-root raised cosine filter with a roll-off factor 0.2 and a bandwidth of $1/T$. The carrier frequency $f_C$ is chosen such that the spectrum of the voiceband is always covered. At the receiver the same square-root raised cosine filter is used for low-pass filtering. Our communication system automatically adjusts the symbol constellation size $Q$, the channel coding rate $R_c$ and the symbol period $T$ such that the best possible data rate is achieved. The receiver knows how well the training symbols were received and based on this feedback the transmitter can optimize the data rate. The relationship between the data rate $R$ and the above parameters is

$$R = \frac{QR_c}{T}. \qquad (4)$$

The receiver used in our setup is iterative. The number of iterations needed for convergence depend on the channel condition. The channel condition is typically measured by means of signal to noise power ratio, the SNR. Figure 6 shows the bit error rate (BER) performance of our receiver for given SNR. The plot also compares two different channel codes. A turbo code (precoded) and a regular convolutional code (not precoded).

Our results are summarized in table 1. The table lists the best compromise between the data rates and the packet drop rate, for various destinations. As can be seen, for our FreeWave server being located in midwest (near Indianapolis) a FreeWave client can get a reliable connection with a bit rate of 19.2 kbps , while a user in Poland achieves a connection bit rate of 16 kbps. These bit rates are convenient for regular web browsing, which is the main intent of censored users.

---

[10]http://software.muzychenko.net/eng/vac.htm

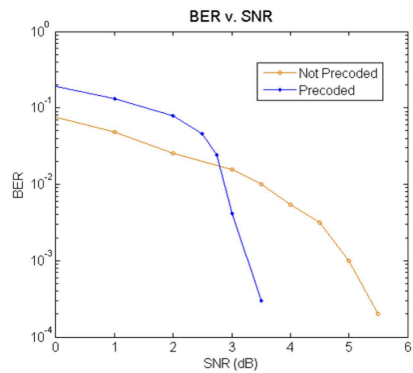

Figure 6: BER versus SNR for FreeWave.

| Data rate ($R$) | $Q$ | $1/T$ | $R_C$ | Packet drop rate | Location of client |
|---|---|---|---|---|---|
| 16000 bps | 4 | 8000 kHz | 0.5 | 0.06 | Lodz, Poland |
| 19200 bps | 4 | 9600 kHz | 0.5 | 0.01 | Chicago, IL |

Table 1: Evaluation results of FreeWave.

Also, note that while FreeWave provides a fairly low bit-rate, as compared to other circumvention systems, its connections is highly unblockable, as discussed throughout this paper.

# 9 Coclusions

In this paper, we presented FreeWave, a censorship circumvention system that is highly unblockable by censors. FreeWave works by modulating a client's Internet traffic inside the acoustic signals that are carried over VoIP connections. Being modulated into acoustic signals, as well as the use of encryption makes FreeWave's VoIP connections unobservable by a censor. By building a prototype implementation of FreeWave we show that FreeWave can be used to achieve connection bit rates that are suitable for normal web browsing.